\documentclass[fleqn,usenatbib]{mnras}

\usepackage{newtxtext,newtxmath}
\usepackage[T1]{fontenc}

\DeclareRobustCommand{\VAN}[3]{#2}
\let\VANthebibliography\thebibliography
\def\thebibliography{\DeclareRobustCommand{\VAN}[3]{##3}\VANthebibliography}

\usepackage{graphicx}	
\usepackage{amsmath}	

\title[EUV waves and a prominence oscillation]{Two successive EUV waves and a transverse oscillation of a quiescent prominence}

\author[Q. M. Zhang et al.]{
Q. M. Zhang,$^{1,2}$\thanks{E-mail: zhangqm@pmo.ac.cn}
M. S. Lin,$^{1,3}$\thanks{E-mail: mslin@pmo.ac.cn}
X. L. Yan,$^{4,2}$
J. Dai,$^{1,5}$
Z. Y. Hou,$^{6}$
Y. Li,$^{1}$
and Y. Qiu$^{7}$
\\
$^{1}$Key Laboratory of Dark Matter and Space Astronomy, Purple Mountain Observatory, Nanjing 210023, China\\
$^{2}$Yunnan Key Laboratory of Solar physics and Space Science, Kunming 650216, China\\
$^{3}$School of Astronomy and Space Science, University of Science and Technology of China, Hefei 230026, China\\
$^{4}$Yunnan Observatories, Chinese Academy of Sciences, Kunming 650216, China\\
$^{5}$Astronomical Observatory, Kyoto University, Sakyo, Kyoto, Japan\\
$^{6}$School of Earth and Space Sciences, Peking University, Beijing 100871, China\\
$^{7}$Institute of Science and Technology for Deep Space Exploration, Suzhou Campus, Nanjing University, Suzhou 215163, China
}

\date{Accepted XXX. Received YYY; in original form ZZZ}

\pubyear{2015}

\begin{document}
\label{firstpage}
\pagerange{\pageref{firstpage}--\pageref{lastpage}}
\maketitle

\begin{abstract}
In this paper, we carry out multiwavelength observations of two successive extreme-ultraviolet (EUV) waves originating from active region (AR) NOAA 13575
and a transverse oscillation of a columnar quiescent prominence on 2024 February 9.
A hot channel eruption generates an X3.4 class flare and the associated full-halo coronal mass ejection (CME), which drives the first EUV wave front (WF1) at a speed of $\sim$835 km s$^{-1}$.
WF1 propagates in the southeast direction and interacts with the prominence, causing an eastward displacement of the prominence immediately.
Then, a second EUV wave front (WF2) is driven by a coronal jet at a speed of $\sim$831 km s$^{-1}$.
WF2 follows WF1 and decelerates from $\sim$788 km s$^{-1}$ to $\sim$603 km s$^{-1}$ before arriving at and touching the prominence.
After reaching the maximum displacement, the prominence turns back and swings for 1$-$3 cycles.
The transverse oscillation of horizontal polarization is most evident in 304 {\AA}.
The initial displacement amplitude, velocity in the plane of the sky, period, and damping time 
fall in the ranges of 12$-$34 Mm, 65$-$143 km s$^{-1}$, 18$-$27 minutes, and 33$-$108 minutes, respectively.
There are strong correlations among the initial amplitude, velocity, period, and height of the prominence.
Surprisingly, the oscillation is also detected in 1600 {\AA}, which is totally in phase with that in 304 {\AA}.
\end{abstract}

\begin{keywords}
Sun: filaments, prominences -- Sun: coronal mass ejections (CMEs) -- Sun: flares -- Sun: oscillations
\end{keywords}



\section{Introduction}
Filaments are cool and dense plasmas, suspending in the solar corona \citep{lab10,mac10,par14}.
According to their positions, filaments are divided into three types, including active region filaments, quiescent filaments, and intermediate filaments \citep{eng98,zou19}.
The latitude of filaments has a wide range from the equator to polar regions \citep{hao15,die24}.
Polar crown filaments (or prominences) are regularly observed in H$\alpha$ and He\,{\sc ii} 304 {\AA} \citep{reg11,tho11,su12}.
The magnetic configuration associated with filaments are mostly sheared arcades and magnetic flux ropes \citep{pri89,au98,dev00,xia11,lr12,yan15}.
\citet{vanb10} proposed that tangled magnetic field in a current sheet is capable of supporting the highly dynamic prominence threads observed in Ca\,{\sc ii} H line \citep{ber08}.

Filaments are prone to rise and erupt when getting unstable, giving rise to solar flares \citep{fle11} and/or coronal mass ejections \citep[CMEs;][]{for06,chen11}.
The impulsive energy releases and expulsions are likely to generate large-scale, propagating waves, 
including coronal extreme-ultraviolet (EUV) waves \citep{tho98,del00,cw11,shen12,liu14,zqm22,zhe22,zhou24} 
and Moreton waves \citep{mor60,uch68,chen02,chen05,eto02,zhe23} at speeds of hundreds of to $\sim$1000 km s$^{-1}$. 
Occasionally, EUV waves associated with type II radio bursts could also be driven by fast coronal jets \citep{su15,mag21,hou23}.

Large-amplitude prominence oscillations, including longitudinal and transverse oscillations, 
are frequently excited and observed by space-borne and ground-based telescopes \citep{tri09,zqm12,arr18,luna18}.
Longitudinal oscillations are usually excited by flares \citep{jing03}, coronal jets \citep{zqm17}, and EUV waves \citep{shen14b}.
Like kink oscillations of coronal loops \citep{asch99,nak99,nak21,guo24}, transverse oscillations of filaments are frequently excited by external disturbances,
such as H$\alpha$ surges \citep{chen08}, coronal jets \citep{zqm17}, EUV waves \citep{her11,lw12,shen14b,shen17,zqm18,devi22,dai23,li24,zyj24}, 
and Moreton waves \citep{eto02,oka04,gil08,asai12,lr13}.
\citet{her11} investigated two successive trains of large-amplitude transverse oscillations of a prominence with a height of $\sim$82 Mm.
The oscillations were induced by two large-scale EUV waves associated with two homologous flares with an interval of $\sim$10 hours on 2005 July 30.
The first wave train had a larger initial amplitude and a shorter damping time than the second one.
\citet{tak15} studied the activation and oscillation of a prominence located at the north pole, 
which was excited by a coronal fast-mode shock wave at a speed of $\sim$670 km s$^{-1}$ on 2012 March 7.
Interestingly, the prominence was strongly compressed and brightened after the arrival of shock wave.

Vertically oscillating filaments are also named winking filaments in spectral observations \citep[e.g.,][]{hyd66,kle69,eto02,shen14a,dai23}.
The filaments appear in the blue wing and red wing of H$\alpha$ line in a staggered way for several cycles before fading out.
Transverse filament oscillations may take place before eruptions \citep{iso06,chen08,dai21,zhou16} or during eruptions \citep{boc11,kum22}.
Hence, transverse oscillations are considered as another precursor of filament eruptions \citep{chen08}.
The amplitudes of oscillations are from a few to tens of Mm, and the periods are from a few to tens of minutes.
The amplitudes usually damp with time as a result of resonant absorption \citep{her11} or wave leakage \citep{kle69}.
The observed periods are employed to estimate the strength of magnetic fields supporting the filaments \citep{hyd66,bal06,shen17,dai23,zyj24}.

In this paper, we report multiwavelength observations of two successive EUV waves and a transverse oscillation of a quiescent prominence 
using the observations of the Atmospheric Imaging Assembly \citep[AIA;][]{lem12} on board the Solar Dynamics Observatory \citep[SDO;][]{pes12} on 2024 February 9.
The paper is organized as follows.
In Section~\ref{data}, we describe the observations of two EUV waves. The results of prominence oscillation are presented in Section~\ref{res}.
Discussions and a brief summary are arranged in Section~\ref{dis} and Section~\ref{sum}, respectively.

\section{Successive EUV waves} \label{data}
\begin{figure}
	\includegraphics[width=\columnwidth]{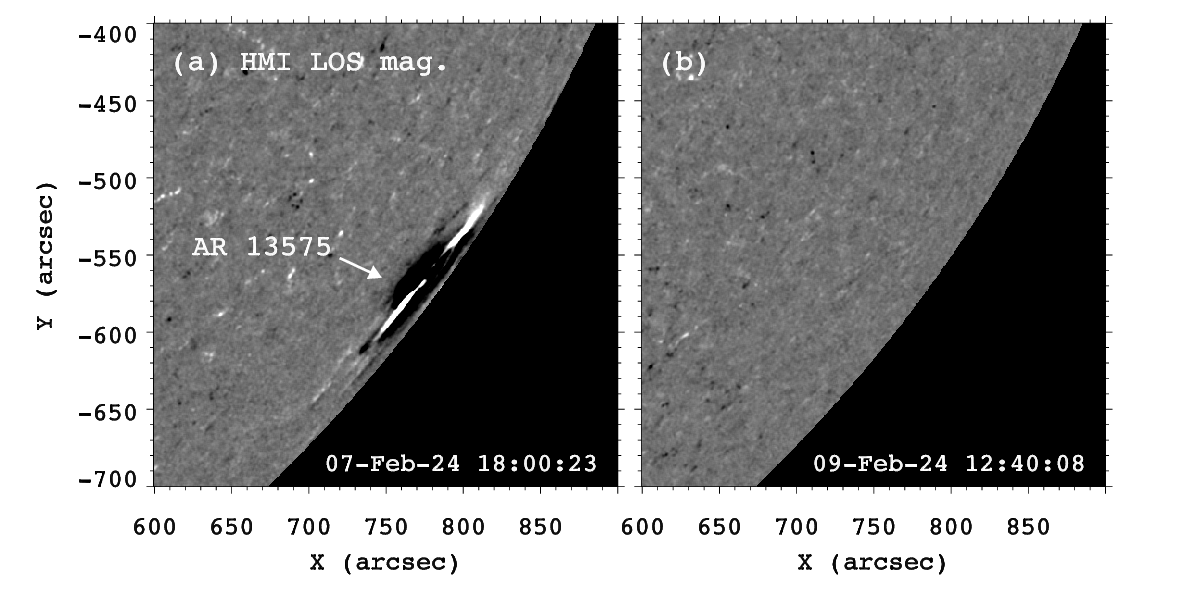}
    \caption{Line-of-sight (LOS) magnetograms observed by SDO/HMI on 2024 February 7 (left panel) and 9 (right panel).
    In panel (a), the white arrow points to AR 13575 close to the western limb.
    In panel (b), the same AR has totally rotated to the far side before flare.}
    \label{fig1}
\end{figure}

\begin{figure}
	\includegraphics[width=\columnwidth]{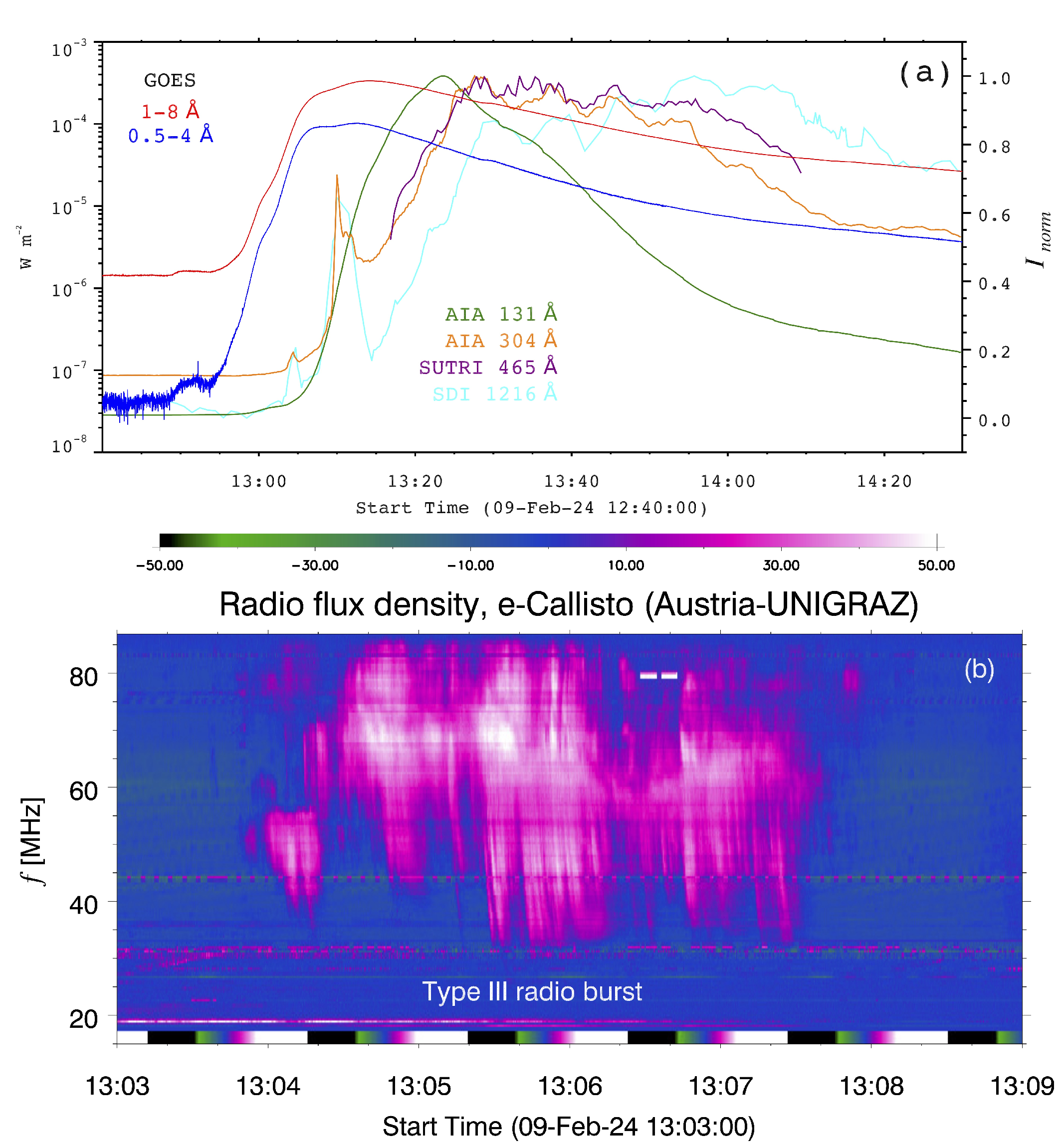}
    \caption{(a) Light curves of the X3.4 class flare in 1$-$8 {\AA} (red line), 0.5$-$4 {\AA} (blue line), 
    131 {\AA} (green line), 304 {\AA} (orange line), 465 {\AA} (purple line), and 1216 {\AA} (cyan line).
    (b) Radio dynamic spectrum observed by the e-Callisto/Austria-UNIGRAZ station during 13:03$-$13:09 UT, featuring a type III radio burst.}
    \label{fig2}
\end{figure}

\begin{figure}
	\includegraphics[width=\columnwidth]{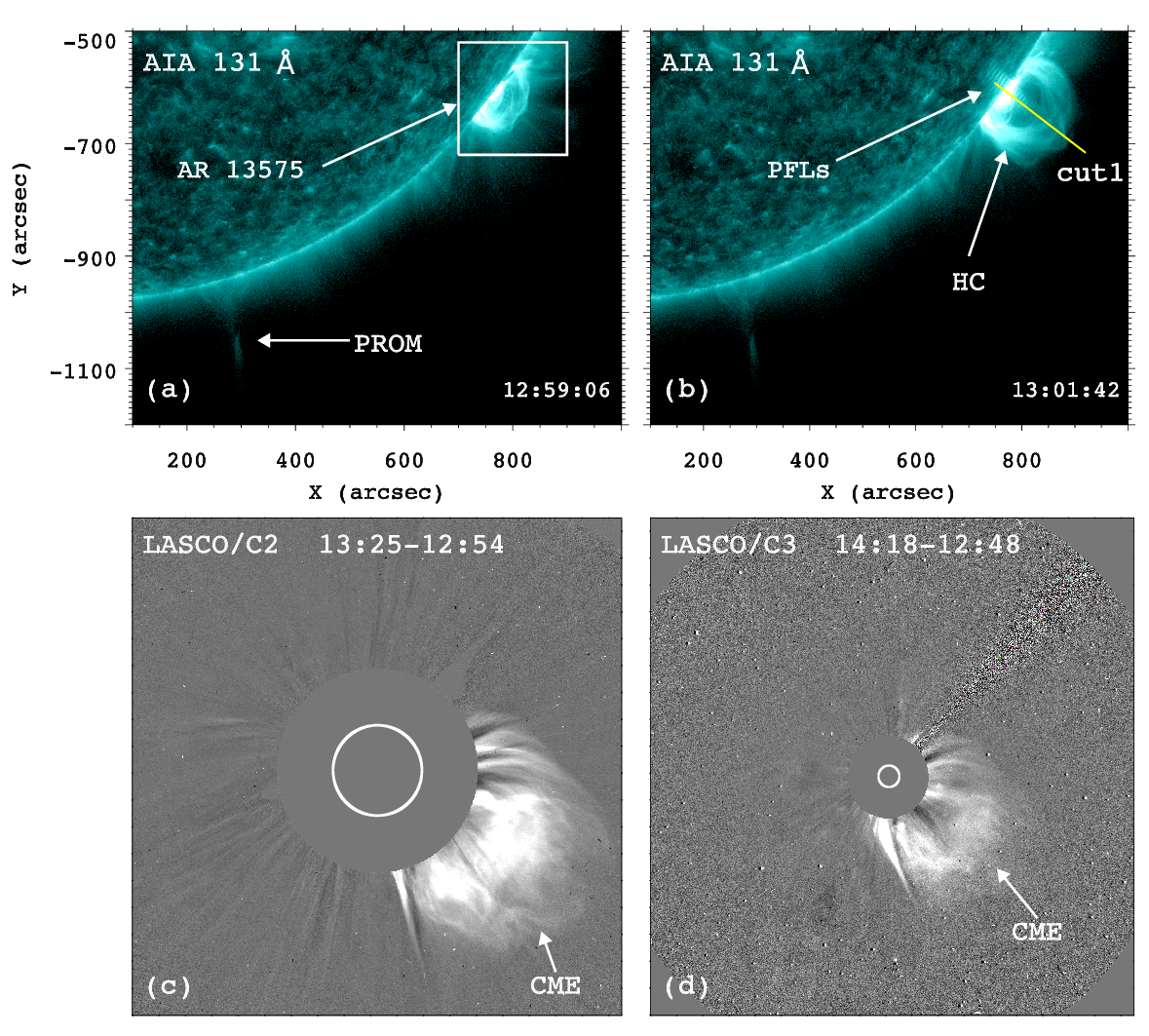}
    \caption{Top panels: SDO/AIA 131 {\AA} images around 13:00 UT, showing the erupting hot channel (HC) and the underlying post-flare loops (PFLs).
    In panel (b), the yellow slice (cut1) is used to investigate the height variation of the HC.
    Bottom panels: running-difference WL images observed by SOHO/LASCO, showing the related full-halo CME.
    An online animation of the AIA 131 {\AA} images is available. The $\sim$9 s animation covers from 12:40 UT to 14:30 UT.}
    \label{fig3}
\end{figure}

\begin{figure}
	\includegraphics[width=\columnwidth]{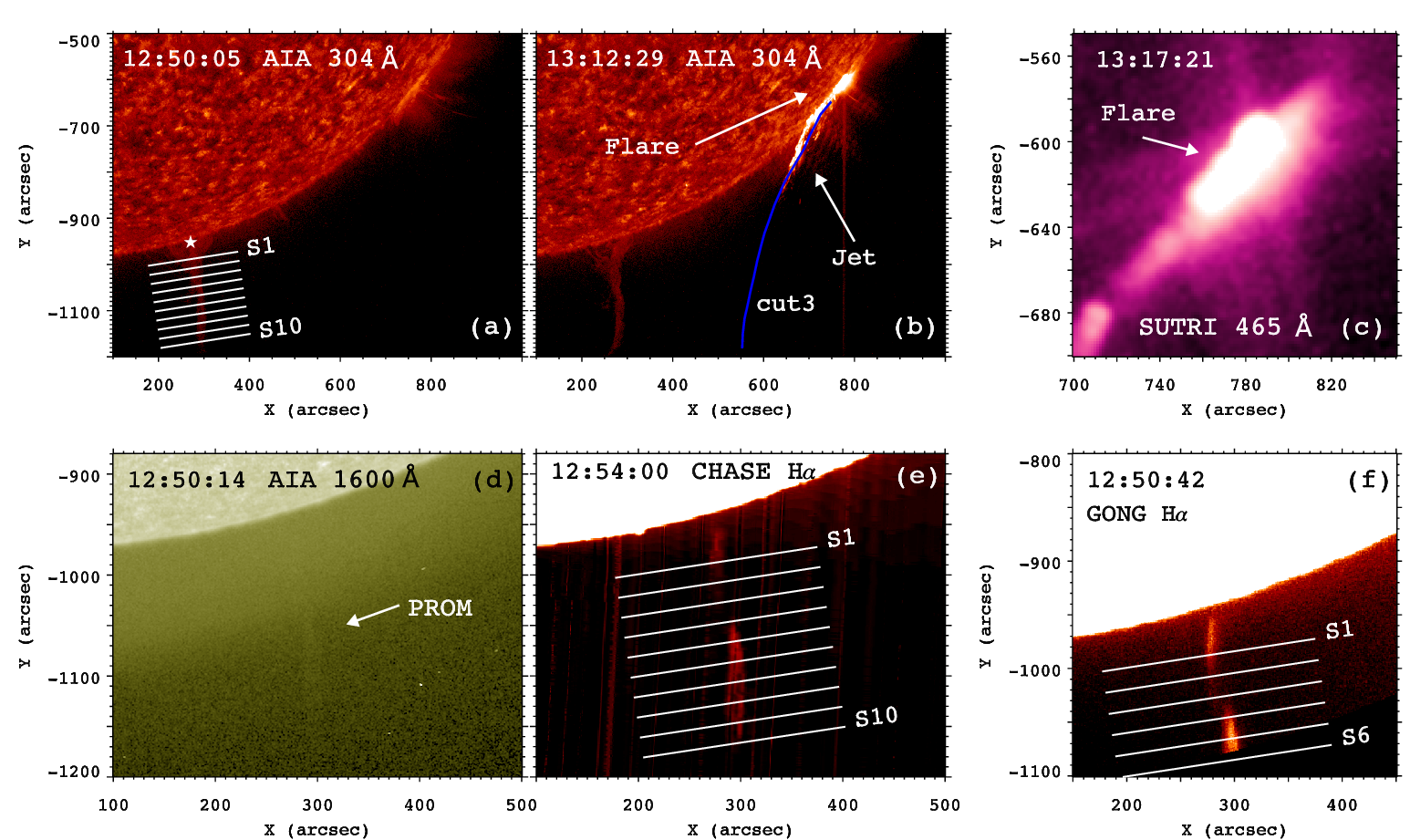}
    \caption{The prominence observed in AIA 304 {\AA} (a-b), 1600 {\AA} (d), CHASE H$\alpha$ (e) and GONG H$\alpha$ (f).
    Ten slices (S1$-$S10) perpendicular to the prominence are used to investigate the transverse prominence oscillation.
    In panel (b), a curved slice (cut3) is used to study the evolution of the coronal jet.
    Panel (c) shows the flare observed in SUTRI 465 {\AA}.
    An online animation of the AIA 304 and 1600 {\AA} images is available. The $\sim$9 s animation covers from 12:40 UT to 14:30 UT.}
    \label{fig4}
\end{figure}

On 2024 February 9, an X3.4 class flare occurred in AR 13575 behind the western limb.
In Figure~\ref{fig1}, the left and right panels show the line-of-sight (LOS) magnetograms observed by the Helioseismic and Magnetic Imager \citep[HMI;][]{sch12} 
on board SDO on February 7 and 9, respectively. It is obvious that AR 13575 (S36W89) was close to the western limb on February 7 and totally rotated to the far side two days later.
In Figure~\ref{fig2}(a), the red and blue lines show SXR light curves of the flare in 1$-$8 {\AA} and 0.5$-$4 {\AA}, respectively.
The SXR flux increases from 12:53 UT, peaks at 13:14 UT, and descends gradually in the decay phase.
In Figure~\ref{fig3}, the top panels show hot post-flare loops (PFLs) of the flare generated by a hot channel \citep[HC;][]{cx11,li13,zqm22,zqm23} eruption originating from AR 13575 
(see also the online movie \textit{anim1.mp4}).
HCs ($T\approx6-10$ MK) are exclusively visible in AIA 94 and 131 {\AA}. It should be emphasized that the footpoints of PFLs are blocked by the western limb.
Close to the south polar region, there is a quiescent prominence, which is $\sim$384 Mm away from the PFLs.
In Figure~\ref{fig4}, the columnar prominence is distinctly observed in 304 {\AA} (panels (a-b)), 1600 {\AA} (panel (d)), and H$\alpha$ line center (panels (e-f)) as well
(see also the online movie \textit{anim2.mp4}).
The H$\alpha$ images were taken by the H$\alpha$ Imaging Spectrograph \citep[HIS;][]{qiu22} on board the Chinese H$\alpha$ Solar Explorer \citep[CHASE;][]{lic22} and GONG, respectively.
The bottom panels of Figure~\ref{fig3} show the consequent full-halo CME at an extremely fast speed of $\sim$2782 km s$^{-1}$ 
when the HC expands and enters into the field of views (FOVs) of C2 and C3 white-light (WL) coronagraphs 
of the Large Angle Spectroscopic Coronagraph \citep[LASCO;][]{bru95} on board the SOHO spacecraft\footnote{cdaw.gsfc.nasa.gov/CME\_list/UNIVERSAL\_ver2/2024\_02/univ2024\_02.html}.

In Figure~\ref{fig3}(a), a white box with a FOV of 200$\arcsec\times$200$\arcsec$ is used to calculate integrated intensities of the flare region in various wavelengths.
The normalized light curves in AIA 131 and 304 {\AA} are plotted with green and orange lines in Figure~\ref{fig2}(a).
The 131 {\AA} emission increases from $\sim$13:00 UT and peaks at $\sim$13:24 UT, which is $\sim$10 minutes later than the SXR peak.
The 304 {\AA} emission reaches the first and second peaks at $\sim$13:05 UT and $\sim$13:10 UT, respectively.
The third peak at $\sim$13:25 UT is followed by a gradual and oscillatory decay.
Fortunately, the flare was detected in H\,{\sc i} 1216 {\AA} with the Solar Disk Imager (SDI) of the Lyman-alpha (Ly$\alpha$) Solar Telescope \citep[LST;][]{li19} 
on board the Advanced Space-based Solar Observatory \citep[ASO-S;][]{gan23} mission.
The corresponding light curve of the flare is plotted with a cyan line in Figure~\ref{fig2}(a).
It is noticed that the first and second peaks in 1216 {\AA} are concurrent with those of 304 {\AA}.
Figure~\ref{fig2}(b) shows the radio dynamic spectrum obtained from the e-Callisto/Austria-UNIGRAZ station\footnote{www.e-callisto.org}.
A type III radio burst, whose frequency rapidly drifts from $\sim$85 to $\sim$30 MHz during 13:04$-$13:08 UT, is the most striking feature.
The dynamic spectrum from another station (EGYPT-Alexandria) shows that the frequency of the radio burst reaches up to $\sim$140 MHz.
This radio burst is roughly coincident with the first peak in both 304 and 1216 {\AA}, implying localized plasma heating in the lower solar atmosphere by flare-accelerated nonthermal electrons.

The flare was also observed with the the Solar Upper Transition Region Imager \citep[SUTRI;][]{bai23} on board the Space Advanced Technology demonstration satellite (SATech-01).
Figure~\ref{fig4}(b-c) show the flare observed by AIA 304 {\AA} at 13:12 UT and by SUTRI Ne\,{\sc vii} 465 {\AA} at 13:17 UT.
Likewise, the normalized light curve in 465 {\AA} during 13:16$-$14:09 UT is derived and drawn with a purple line in Figure~\ref{fig2}(a), which has a very similar variation as in 304 {\AA} (orange line).

\begin{figure}
	\includegraphics[width=\columnwidth]{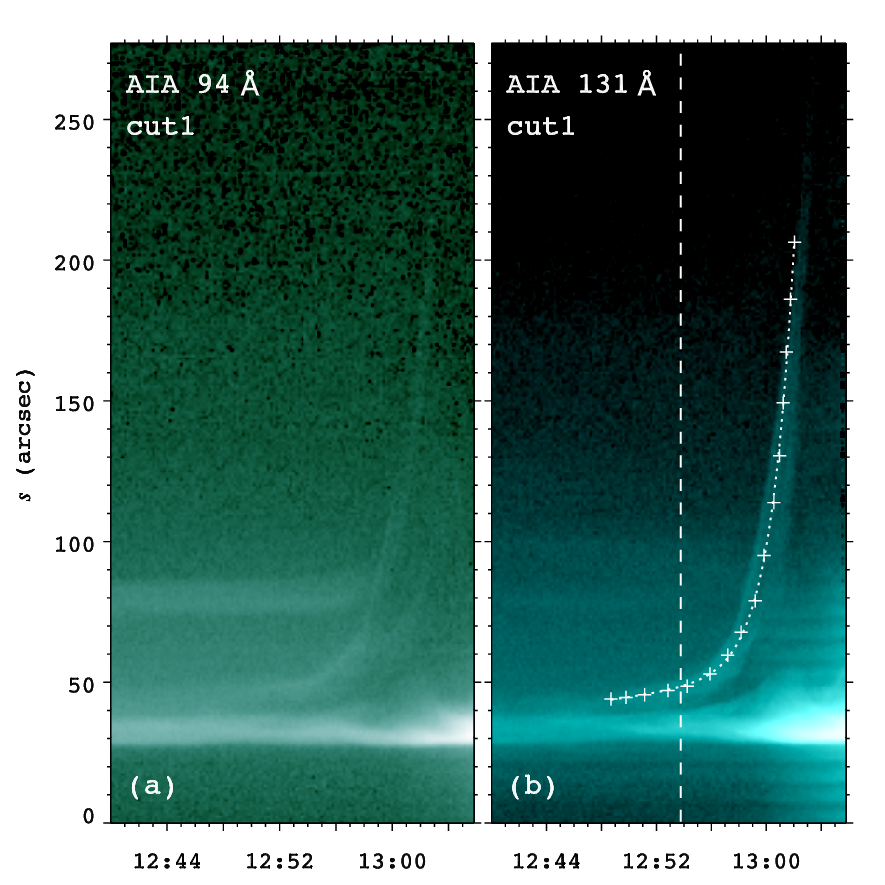}
    \caption{Time-distance maps of cut1 in 94 {\AA} (a) and 131 {\AA} (b).
    $s=0$ and $s=278\arcsec$ denote the northeast and southwest endpoints of cut1, respectively.
    In panel (b), the white ``+'' symbols are trajectory of the HC.
    The dotted line represents the result of curve fitting using Equation~\ref{eqn-1}, and the vertical dashed line represents the onset time of fast rise of the HC.}
    \label{fig5}
\end{figure}

To explore the height variation of the HC, a straight slice (cut1) with a length of $\sim$201 Mm is selected along the direction of eruption in Figure~\ref{fig3}(b).
Time-distance maps of cut1 in 94 and 131 {\AA} are displayed in the left and right panels of Figure~\ref{fig5}.
In Figure~\ref{fig5}(b), the trajectory of HC is marked with white ``+'' symbols.
It is obvious that the HC experiences a slow-rise phase and a fast-rise phase during 12:49$-$13:02 UT, as reported in previous works \citep{cx20,zqm23}.
To fit the trajectory, we use the following equation:
\begin{equation} \label{eqn-1}
  h(t)=c_{0}e^{t/\sigma}+c_{1}t+c_{2},
\end{equation}
where $t$ denotes time after 12:48:39 UT, $h(t)$ is the height of HC, and $\sigma$, $c_{0}$, $c_{1}$, and $c_{2}$ are free parameters.
In Figure~\ref{fig5}(b), the fitted curve is plotted with a dotted line.
The onset time $t_{\mathrm{onset}}$ (12:53:53 UT) of fast rise is marked with a vertical dashed line, which is exactly consistent with the start time of flare impulsive phase.
The final speed of HC reaches $\sim$808 km s$^{-1}$ at 13:02 UT.

\begin{figure}
	\includegraphics[width=\columnwidth]{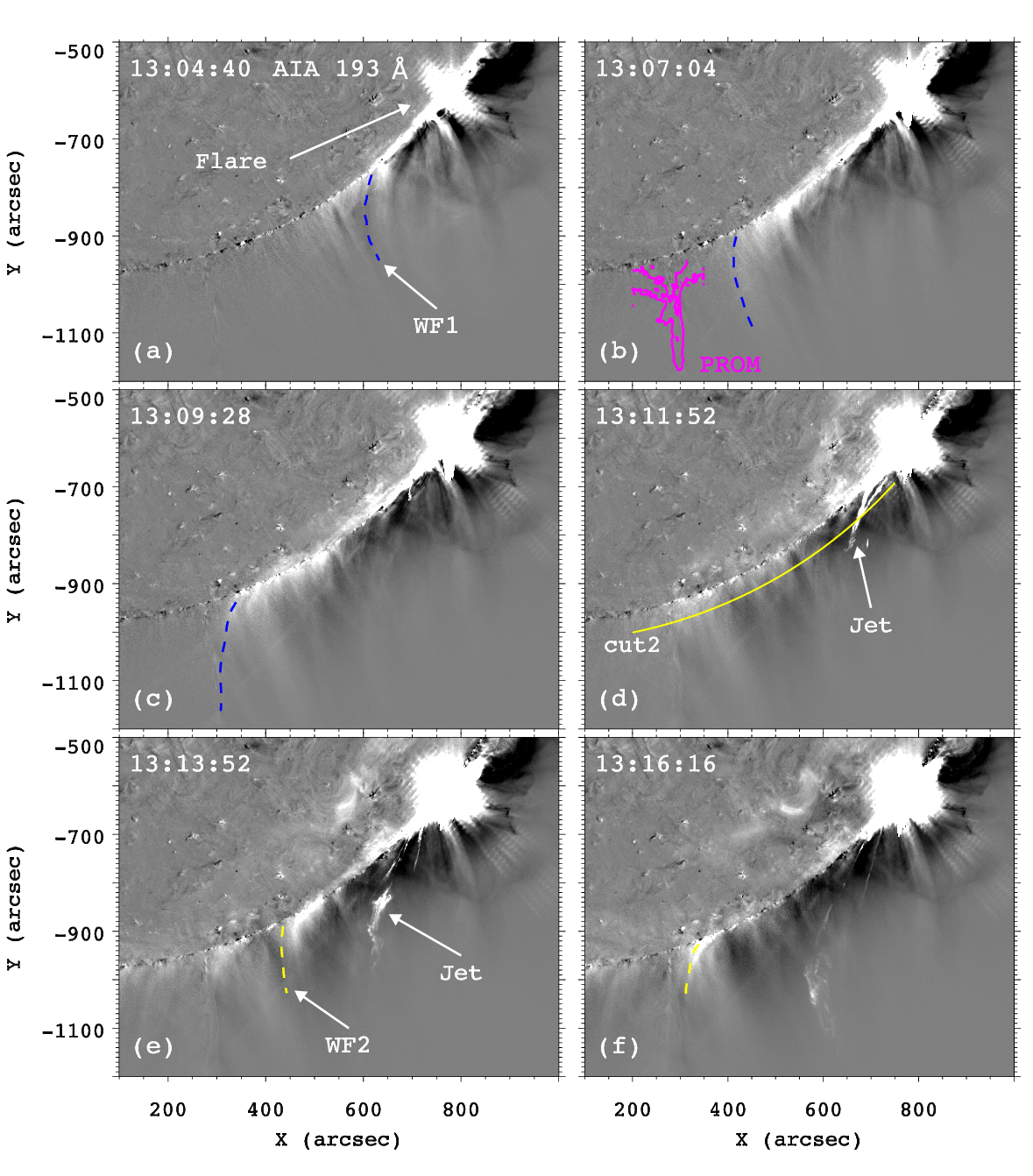}
    \caption{Base-difference images in AIA 193 {\AA} during 13:04$-$13:16 UT.
    The blue dashed lines represent the first EUV wave front (WF1) excited by the HC eruption.
    The yellow dashed lines represent the second EUV wave front (WF2) excited by the coronal jet.
    In panel (b), the magenta lines denote the intensity contours of the prominence in 304 {\AA} at 12:40:05 UT.
    In panel (d), the curved yellow slice (cut2) is used to investigate the evolution of two wave fronts.
    An online animation of the AIA 193 {\AA} base-difference images is available. The $\sim$9 s animation covers from 12:40 UT to 14:30 UT.}
    \label{fig6}
\end{figure}

\begin{figure}
	\includegraphics[width=\columnwidth]{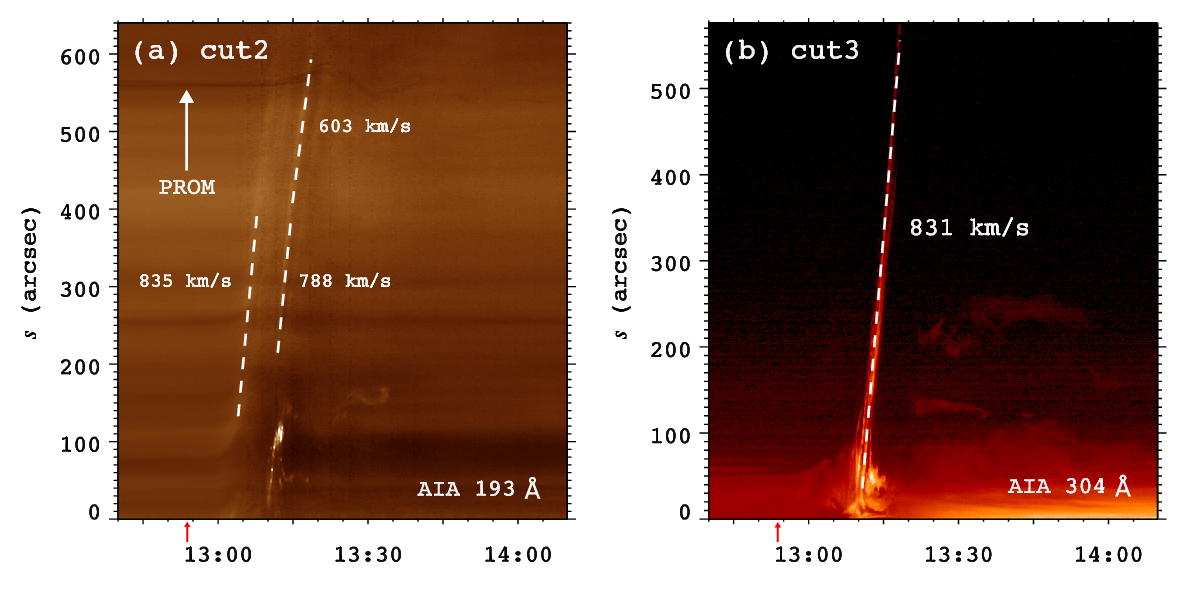}
    \caption{(a) Time-distance map of cut2 in 193 {\AA}. The speeds of two EUV wave fronts are labeled.
    $s=0$ and $s=641\arcsec$ denote the northwest and southeast endpoints of cut2.
    (b) Time-distance map of cut3 in 304 {\AA}. $s=0$ and $s=577\arcsec$ denote the northwest and southeast endpoints of cut3.
    The speed of coronal jet is labeled.}
    \label{fig7}
\end{figure}

The HC eruption and related full-halo CME generate a fast EUV wave propagating in the southeast direction.
Figure~\ref{fig6} shows base-difference images in AIA 193 {\AA} during 13:04$-$13:16 UT (see also the online movie \textit{anim3.mp4}).
As the flare occurs, the first EUV wave front WF1 (blue dashed lines) propagates and sweeps the prominence, 
which is denoted by the intensity contours in 304 {\AA} at 12:40:05 UT (magenta lines) in panel (b).
To calculate the speeds of wave fronts, a curved slice (cut2 with a yellow line) with a length of $\sim$464 Mm and a height of 0.05 $R_{\sun}$ above the solar surface is selected in panel (d).
Time-distance map of cut2 in 193 {\AA} is displayed in the left panel of Figure~\ref{fig7}. It is clear that WF1 starts at $\sim$13:04 UT and propagates at a speed of $\sim$835 km s$^{-1}$.
The WF1 arrives at the prominence at $\sim$13:09 UT and pushes it to move eastward. 
Meanwhile, a coronal jet spurts out from the flare region and propagates in the southeast direction, which is displayed in Figure~\ref{fig4}(b) and Figure~\ref{fig6}(d).
The jet axis has an inclination angle of $\sim$80$^{\circ}$ with the local vertical.
In Figure~\ref{fig4}(b), a curved slice (cut3) with a length of $\sim$418 Mm is selected along the jet axis. Time-distance map of cut3 in 304 {\AA} is displayed in Figure~\ref{fig7}(b).
It is obvious that the jet moves very fast at a speed of $\sim$831 km s$^{-1}$, which is $\sim$1.5 times higher than that of blowout jet on 2022 November 12 \citep{hou23}. 
Interestingly, the jet drives a second EUV wave front WF2 in 193 {\AA}, which is shown with yellow dashed lines in Figure~\ref{fig6}(e-f).
The WF2 decelerates from $\sim$788 km s$^{-1}$ to $\sim$603 km s$^{-1}$ before arriving at the prominence at $\sim$13:17 UT when the prominence is still moving eastwards.
The initial speed of WF2 is slightly higher than that of EUV wave (shock wave) driven by the blowout jet \citep{hou23}.

\begin{table}
	\centering
	\caption{Timeline of the whole events. 
	HC, WF1, WF2 stand for hot channel, the first EUV wave front, and the second EUV wave front, respectively.}
	\label{tab-1}
	\begin{tabular}{cc}
		\hline
		Time (UT) & activity \\
		\hline
		12:49 & Start time of the slow rise of HC \\
		12:53 & Start time of the flare in 1$-$8 {\AA} \\
		12:53 & Onset time of the fast rise of HC \\
		13:04 & Start time of WF1 \\
		13:05 & First peak of the flare in 304 and 1216 {\AA} \\
		13:05 & Type III radio burst \\
		13:09 & WF1 arrives at the prominence \\
		13:09 & Beginning of prominence oscillation \\
		13:10 & Second peak of the flare in 304 and 1216 {\AA} \\
		13:10 & Start time of the coronal jet \\
		13:12 & Start time of WF2 \\
		13:14 & Peak time of the flare in 1$-$8 {\AA} \\
		13:17 & WF2 arrives at the prominence \\
		13:20 & First peak of prominence oscillation \\
		13:25 & CME shows up in LASCO/C2 \\
		14:10 & End time of prominence oscillation \\
		\hline
	\end{tabular}
\end{table}

\section{Prominence oscillation} \label{res}
\begin{figure}
	\includegraphics[width=\columnwidth]{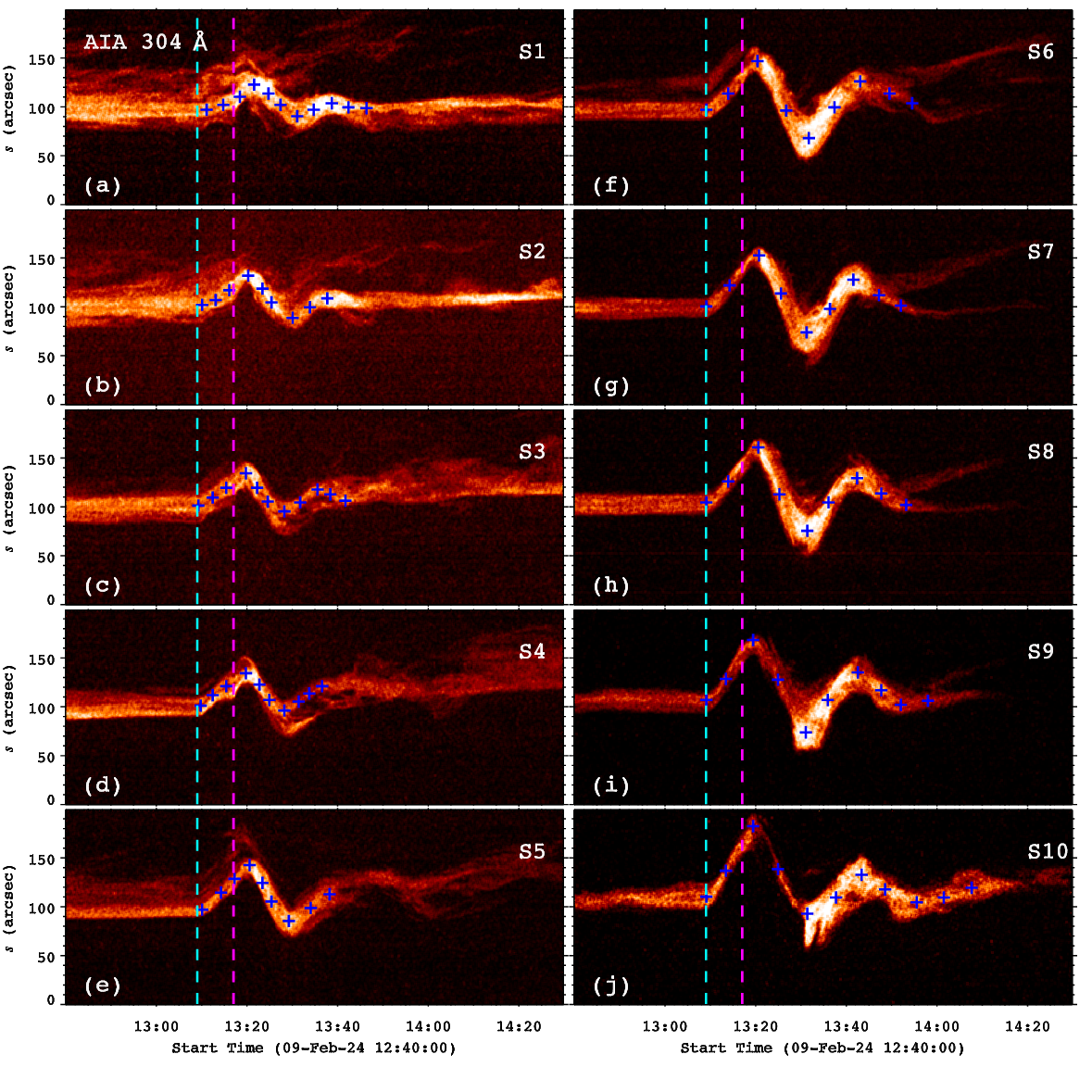}
    \caption{Time-distance maps of S1$-$S10 in 304 {\AA}.
    $s=0$ and $s=200\arcsec$ denote the western and eastern endpoints of the slices, respectively.
    The blue ``+'' symbols represent central positions of the prominence along the slices.
    The cyan and magenta vertical lines represent the arrival times of WF1 and WF2, respectively.}
    \label{fig8}
\end{figure}

\begin{figure}
	\includegraphics[width=\columnwidth]{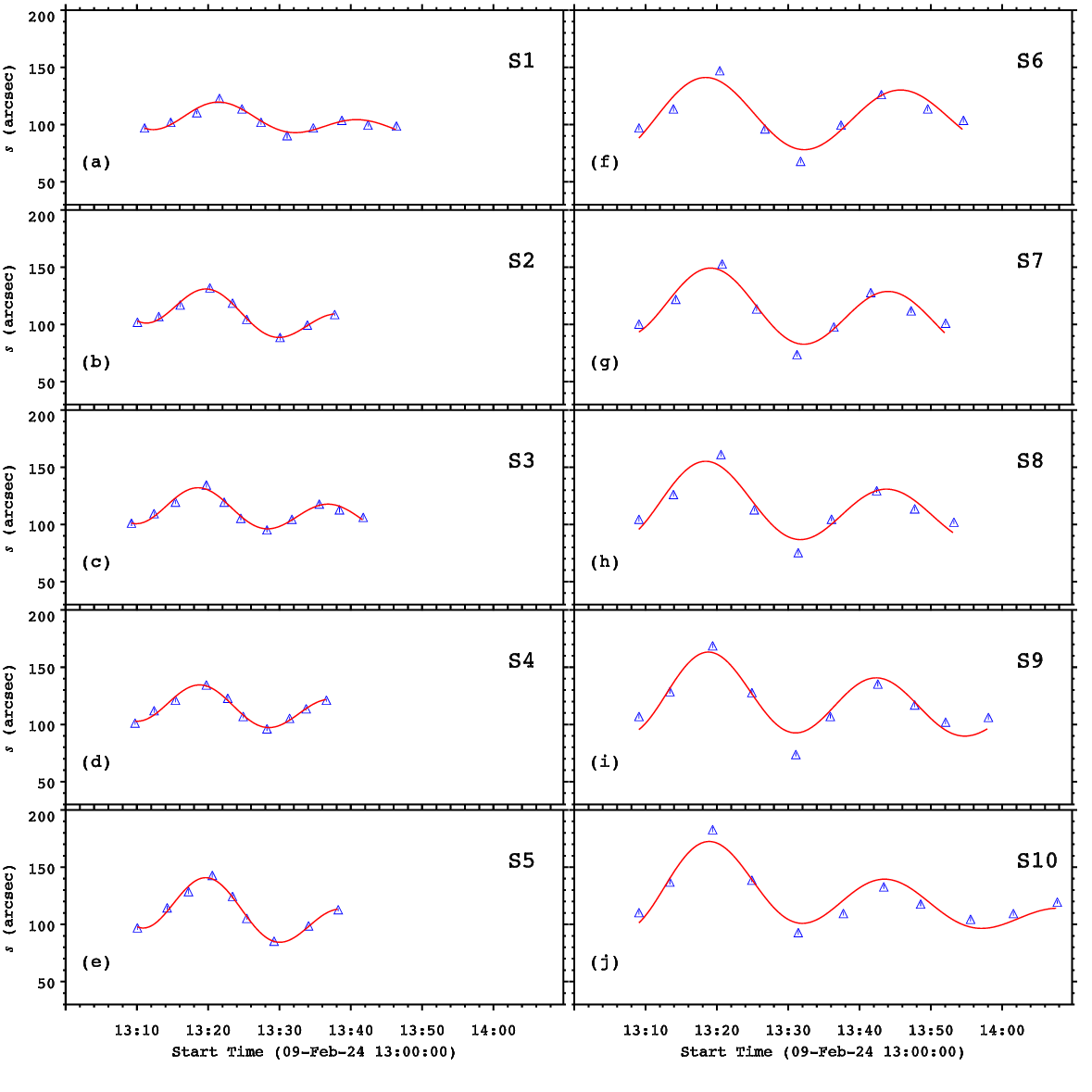}
    \caption{Central positions of the prominence along the ten slices in 304 {\AA} (blue triangles).
    Error bars of the positions are superposed.
    Results of curve fittings using Equation~\ref{eqn-2} are plotted with red lines.}
    \label{fig9}
\end{figure}

As is shown in Figure~\ref{fig3} and Figure~\ref{fig4}, the quiescent prominence with a total length of $\sim$157 Mm is located $\sim$384 Mm away from the PFLs.
The coherent prominence in EUV (131 and 304 {\AA}) wavelengths is seemingly divided into two parts in H$\alpha$, 
which is probably interpreted by a helical magnetic structure of the prominence \citep{cx14}.
As is shown in the online animation \textit{anim3.mp4}, the prominence undergoes a transverse oscillation after the impact of WF1.
In Figure~\ref{fig4}(a), ten slices (S1$-$S10 with the same length of 200$\arcsec$) perpendicular to the prominence are selected to investigate the evolution of prominence.
The heights ($D$) of ten slices increase from 36$\arcsec$ to 216$\arcsec$ (see Table~\ref{tab-2}).
Time-distance maps of these slices in 304 {\AA} are displayed in Figure~\ref{fig8}.
It is clear that the prominence moves eastward after the impact of WF1 at $\sim$13:09 UT (cyan vertical line) 
and the whole body continues to move smoothly after the arrival of WF2 at $\sim$13:17 (magenta vertical line).
It is noted that the impact of WF2 on the prominence is much weaker than that of WF1, 
since only the western side of the prominence is slightly affected when WF2 touches the prominence (Fig. 8(a-d)).
The displacement of the prominence reaches maximum at $\sim$13:20 UT, when it starts to turn back and oscillates horizontally.
The oscillation lasts for 1$-$3 cycles until $\sim$14:20 UT with the amplitude attenuating as time goes on.
To precisely obtain the oscillation parameters, 
the central positions of prominence are marked manually with blue ``+'' symbols in Figure~\ref{fig8} and independently drawn with blue triangles in Figure~\ref{fig9}.
For each slice, the width of the prominence before oscillation is measured, and a quarter of the width is taken to be the error bar of the central position.
The following function is employed to make curve fittings:
\begin{equation} \label{eqn-2}
  y(t)=A_{0}\sin\left[\frac{2\pi}{P}(t-t_{0})+\phi_{0}\right] \mathrm{e}^{-(t-t_{0})/\tau}+y_{0}+k(t-t_{0}),
\end{equation}
where $A_{0}$, $\phi_{0}$, and $y_{0}$ represent the initial displacement amplitude, phase, and position along the slices at $t_0$.
$P$ and $\tau$ represent the period and damping time of the transverse oscillation. $k$ denotes the linear drift speed of the prominence.
The curve fittings are carried out using the standard routine \textsf{mpfit.pro} in the \textit{SolarSoft}\footnote{www.lmsal.com/solarsoft/} packages.
The results of fittings are plotted with red lines in Figure~\ref{fig9}. 
To calculate error bars of the oscillation parameters, we perform 10 Monte Carlo simulations based on error bars of the central positions of the prominence for each slice.
The curve fittings are repeated for 10 times accordingly, and the standard deviations of the parameters are taken to be their error bars.
In Table~\ref{tab-2}, the derived parameters and error bars are listed.

\begin{figure}
	\includegraphics[width=\columnwidth]{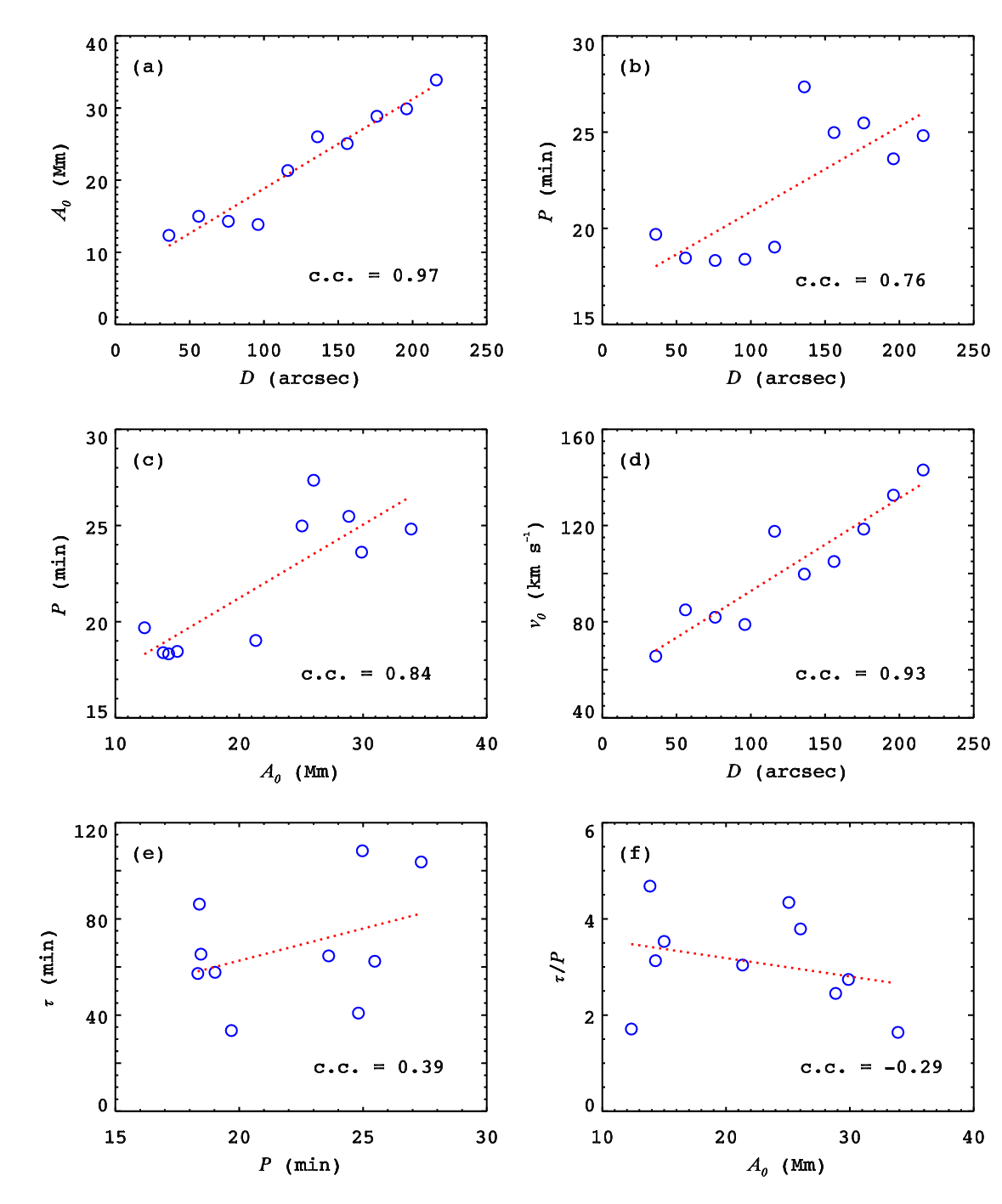}
    \caption{Scatter plots of the parameters of the transverse prominence oscillation.
    $D$ denotes the heights of ten slices.
    $A_0$ and $v_0$ represent the initial displacement amplitude and velocity of oscillation. 
    $P$ and $\tau$ represent the period and damping time, respectively.
    The correlation coefficients (c.c.) are labeled in each panel.}
    \label{fig10}
\end{figure}

Scatter plots of the parameters are displayed in Figure~\ref{fig10}.
The initial amplitude lies in the range of 12$-$34 Mm with an average value of $\sim$22 Mm and roughly increases with the height of the prominence (Figure~\ref{fig10}(a)).
The period lies in the range of 18$-$27 minutes with an average value of $\sim$22 minutes and generally increases with the height (Figure~\ref{fig10}(b)),
which is consistent with the result of transverse oscillations of the prominence on 2005 July 30 \citep{her11}.
The initial amplitude is linearly correlated with the period with a correlation coefficient of $\sim$0.84 (Figure~\ref{fig10}(c)).
The damping time is between 33 and 108 minutes with an average value of $\sim$68 minutes (Figure~\ref{fig10}(e)). 
There is a weak correlation between the period and damping time, i.e., $\tau=2.67P+9.12$.
The quality factor ($q=\tau/P$) is between 1.6 and 4.3 with an average value of $\sim$3.1 (Figure~\ref{fig10}(f)). There is no correlation between the quality factor and period \citep{luna18}.
However, a weak negative correlation between the quality factor and initial amplitude is found, which is similar to the case in kink oscillations of coronal loops \citep{godd16}.
The initial velocity of oscillation ($v_0=2\pi A_0/P$) is between 65 and 143 km s$^{-1}$ with an average value of $\sim$103 km s$^{-1}$ and is linearly correlated with the height (Figure~\ref{fig10}(d)).

\begin{figure}
	\includegraphics[width=\columnwidth]{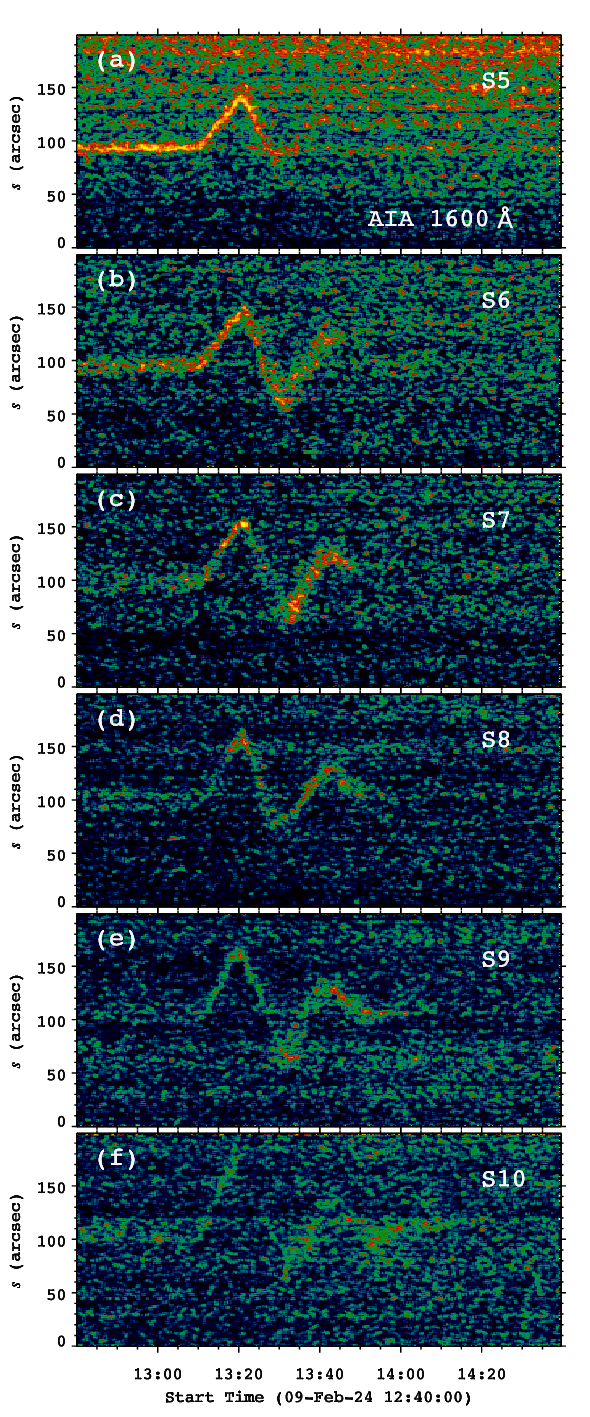}
    \caption{Time-distance maps of S5$-$S10 in UV 1600 {\AA}.}
    \label{fig11}
\end{figure}

Surprisingly, the horizontal prominence oscillation is also detected in AIA 1600 {\AA}, 
although the intensity contrast between the prominence and background is lower than that in 304 {\AA}.
Time-distance maps of S5$-$S10 in 1600 {\AA} are displayed in Figure~\ref{fig11}. 
It is clear that the oscillation across each slice is completely in phase with that in 304 {\AA}.
To our knowledge, this is the first report of prominence oscillation observed in UV 1600 {\AA}.
It is noted that the lower part of the prominence is hard to distinguish from the background. Therefore, the oscillation along S1$-$S4 is quite blurred.
Time line of the whole events is listed in Table~\ref{tab-1}.

\begin{table*}
	\centering
	\caption{Physical parameters of the transverse prominence oscillation along ten slices, 
	including the slice height ($D$), starting time of curve fitting ($t_0$), initial displacement amplitude ($A_0$),
	initial phase ($\phi_0$), period ($P$), damping time ($\tau$), quality factor ($\tau/P$), initial position along the slice ($y_0$),
	linear velocity along the slice ($k$), and initial velocity amplitude ($v_0$).}
	\label{tab-2}
	\begin{tabular}{ccccccccccc}
		\hline
		Slice & $D$ & $t_0$ & $A_0$ & $\phi_0$ & $P$ & $\tau$ & $\frac{\tau}{P}$ & $y_0$ & $k$ & $v_0$ \\
		& [$\arcsec$] & [UT] & [Mm] & [rad] & [min] & [min] & - & [Mm] & [km\,s$^{-1}$] & [km\,s$^{-1}$] \\
		\hline
S1 & 36 & 13:11:04 & 12.35$\pm$0.86 & 4.33$\pm$0.19 & 19.68$\pm$0.77 & 33.57$\pm$5.75 & 1.71$\pm$0.34 & 81.63$\pm$0.92 & -6.10$\pm$0.66 & 65.69$\pm$5.68 \\ 
S2 & 56 & 13:10:04 & 15.00$\pm$0.89 & 4.38$\pm$0.16 & 18.45$\pm$0.55 & 65.32$\pm$13.31 & 3.54$\pm$0.75 & 88.90$\pm$0.93 & -11.46$\pm$0.88 & 85.10$\pm$7.11 \\ 
S3 & 76 & 13:09:14 & 14.30$\pm$1.03 & 4.51$\pm$0.11 & 18.33$\pm$0.41 & 57.28$\pm$13.28 & 3.13$\pm$0.74 & 87.36$\pm$0.65 & -6.47$\pm$0.49 & 81.70$\pm$5.54 \\ 
S4 & 96 & 13:09:44 & 13.85$\pm$0.83 & 4.64$\pm$0.15 & 18.39$\pm$0.65 & 86.12$\pm$14.96 & 4.68$\pm$0.87 & 88.44$\pm$0.81 & -6.07$\pm$0.93 & 78.88$\pm$5.46 \\ 
S5 & 116 & 13:10:04 & 21.33$\pm$1.25 & 4.50$\pm$0.10 & 19.02$\pm$0.41 & 57.77$\pm$6.77 & 3.04$\pm$0.31 & 91.84$\pm$1.33 & -13.06$\pm$1.35 & 117.45$\pm$8.37 \\ 
S6 & 136 & 13:09:04 & 26.01$\pm$1.33 & 5.65$\pm$0.08 & 27.35$\pm$0.48 & 103.64$\pm$20.19 & 3.79$\pm$0.76 & 79.42$\pm$0.67 & -1.49$\pm$0.44 & 99.60$\pm$5.03 \\ 
S7 & 156 & 13:09:04 & 25.07$\pm$0.93 & 5.22$\pm$0.08 & 24.97$\pm$0.33 & 108.29$\pm$19.77 & 4.34$\pm$0.76 & 89.59$\pm$0.74 & -6.77$\pm$0.67 & 105.12$\pm$3.87 \\ 
S8 & 176 & 13:09:04 & 28.85$\pm$0.92 & 5.42$\pm$0.04 & 25.47$\pm$0.27 & 62.36$\pm$4.49 & 2.45$\pm$0.17 & 91.31$\pm$0.81 & -6.14$\pm$0.63 & 118.62$\pm$3.64 \\ 
S9 & 196 & 13:09:04 & 29.88$\pm$0.90 & 5.15$\pm$0.09 & 23.61$\pm$0.40 & 64.60$\pm$6.68 & 2.74$\pm$0.29 & 96.30$\pm$1.24 & -6.01$\pm$0.62 & 132.53$\pm$3.31 \\ 
S10 & 216 & 13:09:04 & 33.88$\pm$1.26 & 5.19$\pm$0.09 & 24.81$\pm$0.51 & 40.82$\pm$2.70 & 1.64$\pm$0.13 & 103.53$\pm$0.96 & -8.03$\pm$0.51 & 143.00$\pm$3.93 \\ 
\hline
Avg.&126 & - & 22.05$\pm$1.02 & 4.90$\pm$0.11 & 22.01$\pm$0.48 & 67.98$\pm$10.79 & 3.10$\pm$0.51 & 89.83$\pm$0.91 & -7.16$\pm$0.72 & 102.77$\pm$5.19 \\		
		\hline
	\end{tabular}
\end{table*}

\section{Discussion} \label{dis}
\citet{kol16} proposed an analytical model of the global transverse oscillations and stability of quiescent prominences. 
In their model, a prominence (magnetic flux rope) is a straight current-carrying wire located at a height $h$ above the photosphere, 
while the magnetic dip is created by two photospheric current sources with a separation of 2$d$.
The periods of horizontal and vertical oscillations with small amplitudes are derived, which closely depend on the parameters of the system.
In a follow-up work, \citet{kol18} investigated the effects of finite amplitudes on the transverse prominence oscillations.
It is found that finite-amplitude horizontal and vertical oscillations are strongly coupled, 
especially for larger amplitudes and smaller attack angles between the direction of the driver (e.g., a shock wave) and the horizontal axis.
In the nonlinear large-amplitude regime, the horizontal period ($P_x$) always increases with the horizontal amplitude (see top panels of their Fig. 10).
For a fixed amplitude, the horizontal period increases with the prominence height as well.
In our study, the scatter plots in Figure~\ref{fig10} indicate that the period of transverse oscillation increases with the prominence height (panel (b)) and the displacement amplitude (panel (c)).
In this sense, the large-amplitude, horizontal prominence oscillation excited by the EUV waves on 2024 February 9 could be qualitatively explained by the model of \citet{kol18}.
Sophisticated MHD numerical simulations are worthwhile to reproduce and explain the observational results thoroughly \citep{li23}.
 
Successive coronal loop oscillations induced by homologous flares have been observed before \citep{nis13,zqm20}.
In our study, the prominence is impacted by two successive EUV wave fronts, WF1 driven by the CME and WF2 driven by the jet, which have a short interval of $\sim$8 minutes (Table~\ref{tab-1}).
However, the prominence has not returned back when WF2 arrives. Acturally, the prominence is slightly touched by WF2 and continues to move eastward.
Therefore, the two wave fronts push the prominence in the same direction, rather than opposite directions.
Consequently, the prominence undergoes a single oscillatory motion.

Successive EUV waves in a single eruption event have been observed. \citet{zhe22} investigated twin EUV waves in the solar corona.
In the first case, the two waves are separately driven by a filament eruption and a precursor jet, i.e., the jet occurs prior to the filament eruption.
Linear speeds of the wave trains are $\sim$230 and $\sim$390 km s$^{-1}$.
In the second case, the two waves are successively associated with a filament eruption.
The two EUV wave fronts in our study are similar to their first case. 
The difference is that the EUV wave fronts are primarily driven by the CME as a result of a HC eruption and then driven by the coronal jet, which has the opposite order to their case.
Besides, the EUV waves in our study are fast-mode waves at speeds of $\sim$835 and $\sim$788 km s$^{-1}$, which are significantly faster than the wave trains studied by \citet{zhe22}.

\section{Summary} \label{sum}
In this paper, we carry out multiwavelength observations of two successive EUV waves originating from AR 13575
and a transverse oscillation of a columnar quiescent prominence with a total length of $\sim$157 Mm on 2024 February 9.
The main results are summarized as follows:
\begin{enumerate}
   \item The first EUV wave front (WF1) is driven by a CME as a result of a HC eruption, which also generates an X3.4 class flare. The speed of WF1 reaches $\sim$835 km s$^{-1}$.
   After the impact of WF1, the prominence moves eastward immediately.
   Then, a second EUV wave front (WF2) is driven by a coronal jet at a speed of $\sim$831 km s$^{-1}$.
   WF2 follows WF1 and decelerates from $\sim$788 km s$^{-1}$ to $\sim$603 km s$^{-1}$ before arriving at and touching the prominence.
   After reaching the maximum displacement, the prominence turns back and swings for 1$-$3 cycles.
   The horizontally polarized oscillation is most striking in AIA 304 {\AA} and is surprisingly detected in 1600 {\AA}, which are in phase in the two wavelengths.
   \item The initial displacement amplitude, velocity in the plane of the sky, period, and damping time 
   fall in the ranges of 12$-$34 Mm, 65$-$143 km s$^{-1}$, 18$-$27 minutes, and 33$-$108 minutes, respectively.
   There are strong correlations among the initial amplitude, velocity, period, and height of the prominence.
\end{enumerate}

\section*{Acknowledgements}
The authors appreciate the reviewer for valuable comments and suggestions to improve the quality of this article.
We also appreciate Prof. Liheng Yang in Yunnan Astronomical Observatories for helpful discussions.
SDO is a mission of NASA\rq{}s Living With a Star Program. AIA and HMI data are courtesy of the NASA/SDO science teams.
The CHASE mission is supported by China National Space Administration (CNSA).
SUTRI is a collaborative project conducted by the National Astronomical Observatories of CAS, Peking University, Tongji University, 
Xi'an Institute of Optics and Precision Mechanics of CAS, and the Innovation Academy for Microsatellites of CAS.
The ASO-S is supported by the Strategic Priority Research Program on Space Science, Chinese Academy of Sciences.
This work is supported by the Strategic Priority Research Program of the Chinese Academy of Sciences, Grant No. XDB0560000,
the National Key R\&D Program of China 2021YFA1600500 (2021YFA1600502), 2022YFF0503003 (2022YFF0503000), 
NSFC under the grant numbers 12373065, 12325303, 12273115, 12303057, Natural Science Foundation of Jiangsu Province (BK20231510), 
and Yunnan Key Laboratory of Solar Physics and Space Science under the grant numbers 202205AG070009, YNSPCC202206.
J. Dai is supported by the Special Research Assistant Project of Chinese Academy of Sciences, 
the Project funded by China Postdoctoral Science Foundation (2023M733734).

\section*{Data Availability}
The data underlying this article will be shared on reasonable request to the corresponding author.

\bibliographystyle{mnras}
\bibliography{po}                                                          

\bsp	
\label{lastpage}
\end{document}